\documentclass[aps,floatfix,superscriptaddress,notitlepage]{revtex4-1}
\usepackage{amsmath,amssymb,amsfonts,graphics,graphicx,dcolumn,bm,enumerate}
\usepackage{comment,natbib,appendix}
\usepackage{multirow,color}
\usepackage{chngpage}

\usepackage{afterpage}
\usepackage{xcolor}
\usepackage{amsthm}
\usepackage{epsfig}
\usepackage{natbib}
\usepackage{hyperref}
\usepackage[margin=1.2in]{geometry}
\usepackage{graphicx}

\newcommand{\cmp}
{\affiliation{Saha Institute of Nuclear Physics, Kolkata 700064, India.}}
\newcommand{\isi}
{\affiliation{Economic Research Unit, Indian Statistical Institute, Kolkata 700108, India.}}
\newcommand{\raghunathpur}
{\affiliation{Raghunathpur College, Raghunathpur, Purulia 723133, India.}}
\newcommand{\snb}
{\affiliation{S. N. Bose National Centre for Basic Sciences, Kolkata 700106, India}}

\begin{document}

 \title{Limiting Value of the Kolkata Index for Social
Inequality and a Possible Social Constant}

 \author{Asim Ghosh}
 \email[Email: ]{asimghosh066@gmail.com}
 \raghunathpur
 \author{Bikas K Chakrabarti}%
 \email[Email: ]{bikask.chakrabarti@saha.ac.in}
 \cmp \isi \snb

\begin{abstract}
\noindent Based on some analytic structural properties
of the Gini and Kolkata indices for social
inequality, as obtained from a generic form of
the Lorenz function, we make a conjecture that
the limiting (effective saturation) value
of the above-mentioned indices is about 0.865.
This, together with some more new observations
on the citation statistics of individual
authors (including Nobel laureates), suggests
that about $14\%$ of people or papers or social
conflicts tend to earn or attract or cause
about $86\%$ of wealth or citations or deaths
respectively in very competitive situations
in markets, universities or wars. This is a modified form of the (more
than a) century old $80-20$ law of Pareto in
economy (not visible today because of various
welfare and other strategies) and  gives an
universal value ($0.86$) of social (inequality)
constant or number.

\end{abstract}

\maketitle

\section{Introduction}
\noindent Unlike the universal constants in physical
sciences, like the Gravitational Constant
of Newton's Gravity law, Boltzmann Constant
of thermodynamics or Planck's Constant of
Quantum Mechanics, there is no established
universal constant yet in social sciences.
There have of course been suggestion of
several possible candidates.

Stanley Milgram's experiment \cite{Milgram1967small} to determine
the social `contact-distance' between  any two
persons of the society, by trying to deliver
letters from and to random people through
personal chains of friends or acquaintances,
suggested `Six Degrees of Separation'. Studying similar distance through
co-authorship of papers, between any two
scientists (e.g., the Erdos number \cite{wiki},
describing the collaborative distance
between mathematician Paul Erdos and
another mathematician) indicated similar
but not identical numbers.   Later, the (internet) network structure
studies \cite{Newman2006Structure,Barabasi2014Everything} linked the (separation) number
to be related to the network size (typically
going as log of the network size) and not 
really as universal as six. The Dunbar number \cite{Dunbar1992Neocortex}, suggesting that
we can only maintain one hundred and fifty
distinct  social relationships (as may be
seen in the sizes of the old village groups),
has  also been questioned. It is observed
to vary from much smaller numbers, for closer
shell relationships, to order of magnitude
larger number, for social weblinks and can be
extracted, say, from the sizes of individual's
mobile call list (see e.g., \cite{McCarty2000Comparing,Bhattacharya2016sex}).

We find that the limiting magnitude  of a
particular social inequality measure shows a
robust and universal value across different
social contexts. In a series of papers \cite{Ghosh2014Inequality,Ghosh2016kinetic,Chatterjee2017Socio-economic,Sinha2019Inequality,Banerjee2020Kolkata1}
(see also \cite{Banerjee2020Kolkata2} for a recent review), we
introduced the Kolkata index ($k$) for measuring
social inequality ($k = 1/2$ corresponds to
perfect equality and $k = 1$ corresponds to
extreme inequality). In the economic context \cite{Ghosh2014Inequality},
it says $(1 - k)$ fraction of people posses $k$
fraction of wealth, while in the context of an
university \cite{Ghosh2014Inequality,Chatterjee2017Socio-economic}, it says $(1 - k)$ fraction of
papers published by the faculty of the university
attracts $k$ fraction of citations, or even in the
context of wars or major social conflicts, it
says \cite{Sinha2019Inequality} $(1 - k)$ fraction of social conflicts
or wars cause $k$ fraction of deaths. We observed \cite{Ghosh2014Inequality,Chatterjee2017Socio-economic,Sinha2019Inequality}, in a very wide
range of social contexts, the limiting (or
effective saturation) value of the Kolkata
index $k$ to be around $0.86$ (except in the
case of world economies today, where such
limiting value of $k$ is observed to be
much smaller and is about $0.73$).
Indeed, the $k$ index is a quantitative
generalization of the century old $80-20$ rule of
Vilfredo Pareto \cite{Pareto1971Translation}, who observed towards the
end of eighteenth century that in most of the
European countries (Italy, in particular)
almost $80\%$ of the land are owned by $20\%$ of
the people (i.e., $k \sim 0.80$), and perhaps
similar  across the  entire economy in those days
(when massive economic welfare measures or land
reforms, etc. did not exist!).

%%%%%%%%%%%%%%%%%%%%%%%%%%%%%%%%%%%%%%%%%%%%%%%%%%%%%%%%%% FIG1
\begin{figure}[!h]
\centering
\includegraphics[width=10cm]{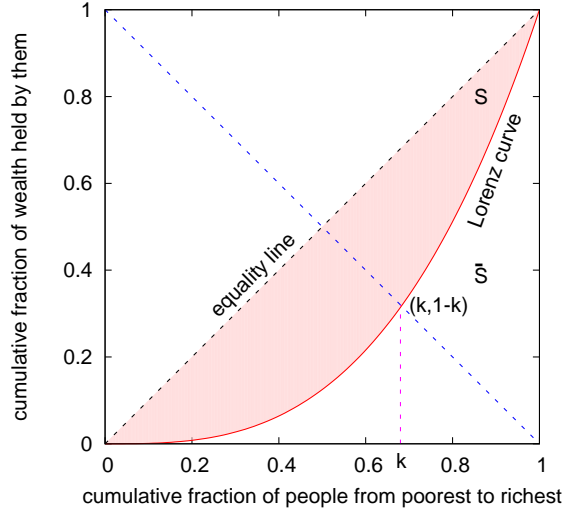}
 \caption{Lorenz curve or function $L(x)$ (in red here)
represents the accumulated fraction of wealth
(or citations or deaths) against the fraction
$(x)$ of people (or papers or social conflicts)
possessing (or attracting or causing) that, when
arranged from poorest (or lowest) to richest
(or highest). The diagonal from the origin
represents the equality line. The Gini index
$(g)$ can be measured \cite{Gini1921Measurement} by the area $(S)$
of the shaded region in-between the Lorenz
curve and the equality line, when normalized by
the area $(S + \bar S = 1/2)$ of the triangle
below the equality line): $g = 2S$.  The Kolkata
index $k$ can be measured by the ordinate value
of the intersecting point of the Lorenz curve and
the diagonal perpendicular to the equality line.
By construction, it says that $k$ fraction of
wealth (or citations or deaths) is being held by
$(1 - k)$ fraction of top people (or papers or
social conflicts).}
 \label{Lorentz2.pdf}
\end{figure}
%%%%%%%%%%%%%%%%%%%%%%%%%%%%%%%%%%%%%%%%%%%%%%%%%%%%%%%%%%

We will first discuss here analytically some
indications of a limiting behavior of the Kolkata
index, suggesting its value $k$ near 0.86. Next
we will provide some  detailed analysis of data
from different social sectors like citations of
papers published by different universities and in
different journals, human deaths in different wars
or social conflicts, and citations of papers
published by individual authors (including Nobel
Laureates) showing  that the limiting value of
the inequality index $k$ suggests that typically
$86\%$ of citations (or deaths) come from $14\%$
papers (or conflicts).

\section{LORENZ CURVE: GINI \& KOLKATA INDICES}
\noindent Our study here  is based on the Lorenz curve (see
Fig. \ref{Lorentz2.pdf})  or function \cite{Lorenz1905Methods} $L(x)$, which  gives
the cumulative fraction of (total accumulated)
wealth (or citations or human deaths) possessed
(attracted or caused) by the fraction $(x)$ of
the people (or papers or social conflicts
respectively) when  counted from the poorest
(or least or mildest) to the richest (or
highest or deadliest). If the income/wealth
(or citations or deaths) of every person (or
paper or war) would be identical, then $L(x)$
would be a straight line (diagonal) passing
through the origin. This diagonal is called
the equality line. The Gini coefficient or index $(g)$ is
given by twice the area between the Lorenz
curve and the equality line: $g= 0$
corresponds to equality and $g= 1$
corresponds to extreme inequality.

%%%%%%%%%%%%%%%%%%%%%%%%%%%%%%%%%%%%%%%%%%%%%%%%%%%%%%%%%% fig2
\begin{figure}[!h]
\centering
\includegraphics[width=10cm]{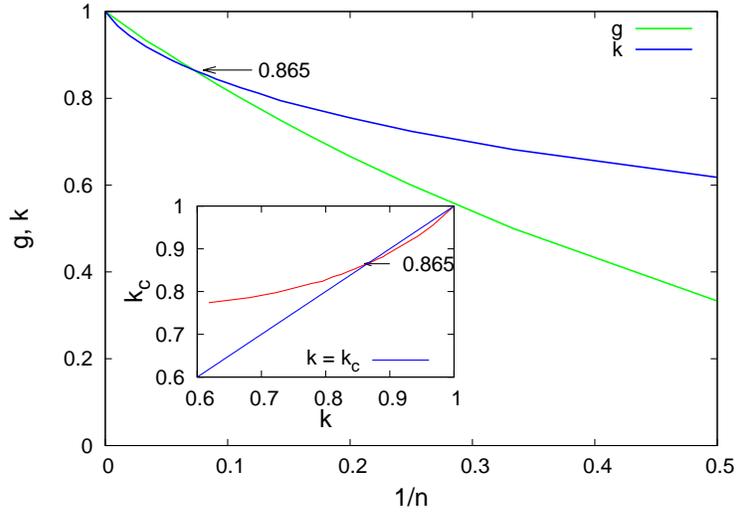}
 \caption{Gini $(g)$ and Kolkata $(k)$ indices obtained
numerically for the generic form of the Lorenz
function $L(x) = x^n$ (eqn. (\ref{eqn1}); $n$ is a
positive real number) for different values of $1/n$.
For $n$ = 1, $g = 0$ and $k = 0.5$ and as
$n \to \infty$ $g = 1 = k$. However, at $g \simeq 0.865 \simeq k$, $g$
crosses $k$ and then turns again and become
equal at extreme inequality. This multi-
valued equality property of $k/g$ as function
of $k$ seems to restrict the inequality
measure at the limiting value of $k(=g)$ at
about 0.865 below its extreme value at unity. This  multi-valued equality property
of $k$ as function  of $g$ seems to restrict
the inequality measure at the limiting value of
 $k$ (= $g$) at 0.86 below its extreme value at
unity. The inset shows  the $k_c$ values
obtained by fitting the  different $k$ and $g$
values (for different $n$) to the linear equation
(\ref{eqn2}) and then solving for $k = k_c = g$.}
 \label{fig2:01-03-20-n-vs-gk.eps}
\end{figure}
%%%%%%%%%%%%%%%%%%%%%%%%%%%%%%%%%%%%%%%%%%%%%%%%%%%%%%

%%%%%%%%%%%%%%%%%%%%%%%%%%%%%%%%%%%%%%%%%%%%%%%%%%%%%%%%%% fig3 
\begin{figure}[!h]
\centering
\includegraphics[width=10cm]{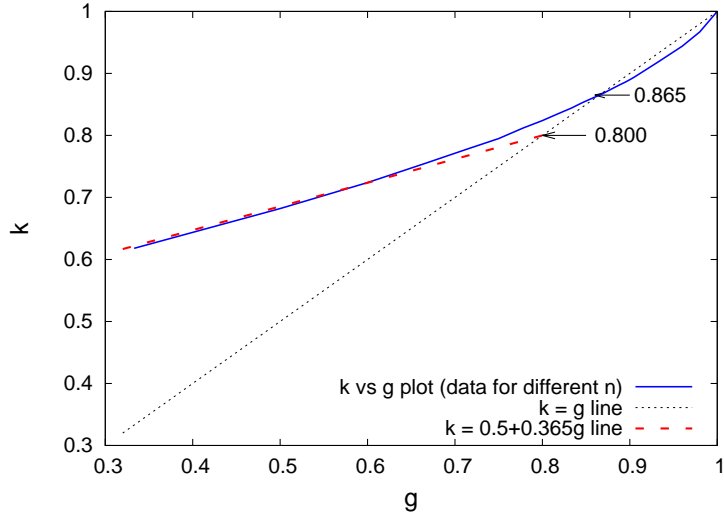}
 \caption{Shows the plot of the different $k$ values vs.
corresponding $g$ values (for different $n$
in eqn. (\ref{eqn1}); see Fig. \ref{fig2:01-03-20-n-vs-gk.eps}). Also the $k = g$ line
is shown. A linear extrapolation (\ref{eqn2}) of the Initial
part of $k$ vs. $g$ suggests $k=g= 0.800$, while
they really become equal following eqn. (\ref{eqn1}) at
$k=g \simeq 0.865$.}
 \label{fig3:01-03-20-k-vs-g.eps}
\end{figure}
%%%%%%%%%%%%%%%%%%%%%%%%%%%%%%%%%%%%%%%%%%%%%%%%%%%%%%%%%%

We proposed \cite{Ghosh2014Inequality} the Kolkata index ($k$) given
by the ordinate value (see Fig. \ref{Lorentz2.pdf}) of the
intersecting point of the Lorenz curve and the
diagonal perpendicular to the equality line
(see also \cite{Ghosh2016kinetic,Chatterjee2017Socio-economic,Sinha2019Inequality,Banerjee2020Kolkata1,Banerjee2020Kolkata2}). By construction, $1 - L(k)$
$= k$, saying that $k$  fraction of wealth (or
citations or deaths) is being possessed (owned
or caused)  by $(1- k)$ fraction of the richest
population (impactful papers or deadliest wars).
As such, it gives a quantitative generalization
of more than a century old phenomenologically
established 80-20 law of Pareto \cite{Pareto1971Translation}, saying
that in any economy, typically about 80\% wealth
is possessed by only 20\% of the richest
population. Defining the Complementary Lorenz
function $L^{(c)}(x) \equiv [1 - L(x)]$, one gets $k$
as its (nontrivial) fixed point \cite{Banerjee2020Kolkata1,Banerjee2020Kolkata2}:
$L^{(c)}(k) = k$ (while Lorenz function $L(x)$
itself has trivial fixed points at $x = 0$ and
1). Kolkata index ($k$) can also be viewed as a
normalized  Hirsch index ($h$)\cite{Lorenz1905Methods} for social
inequality as $h$-index is given by the fixed
point value of the nonlinear citation function
against the number of publications of individual
researchers. We have studied the mathematical
structure of $k$-index in \cite{Banerjee2020Kolkata1} (see \cite{Banerjee2020Kolkata2} for a
recent review) and its suitability, compared
with the Gini and other inequality indices or
measures, in the context of different social
statistics, in \cite{Ghosh2014Inequality,Ghosh2016kinetic,Chatterjee2017Socio-economic,Sinha2019Inequality,Banerjee2020Kolkata1,Banerjee2020Kolkata2}.

\section{Numerical study of $g$ and $k$ for a generic from of Lorenz function}
\noindent For various distributions of wealth, citations or
deaths, the generic form of the Lorenz function
$L(x)$  is such that $L(0) = 0$ and $L(1) = 1$
and it grows monotonically with $x$. As a generic
form, we assume

\begin{equation}
 L(x) = x^n,
 \label{eqn1}
\end{equation}
\noindent where $n$ is a positive real number. For $n = 1$, the
Lorenz curve falls on the equality line and one gets
$g = 0$, $k=0.5$. For $n = 2$, $g = 1/3$  and $k
= ((\sqrt 5 - 1)/2)$ becomes inverse
of the Golden ratio \cite{Ghosh2016kinetic,Banerjee2020Kolkata1,Banerjee2020Kolkata2}. For increasing values
of $n$,  both $g$  and $k$ approach unity, and our
numerical study indicates some interesting 
non-monotonic variational relationship between $g$ and
$k$ (see also \cite{Ghosh2016kinetic} for similar features in the case of special distributions),
and shown in Figs. \ref{fig2:01-03-20-n-vs-gk.eps} and \ref{fig3:01-03-20-k-vs-g.eps}.

%%%%%%%%%%%%%%%%%%%%%%%%%%%%%%%%%%%%%%%%%%%%%%%%%%%%%%%%%% fig4
\begin{figure}[!h]
\centering
\includegraphics[width=10cm]{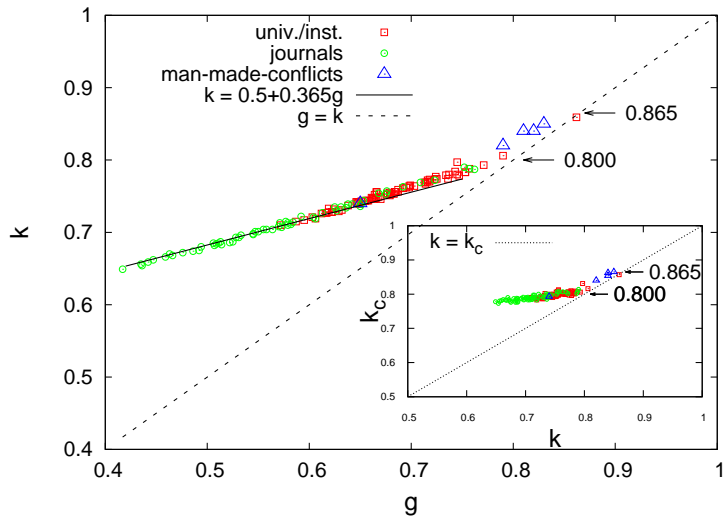}
 \caption{Plot of estimated values of $k$  against $g$
from the the web of science data for citations
against papers published by authors from
different universities or institutes  and also
of the publications in different journals (from
refs. \cite{Ghosh2014Inequality,Chatterjee2017Socio-economic}).  Similar data for the death
distributions in various social conflicts or
wars \cite{Sinha2019Inequality} are also shown. The inset shows
the linear extrapolation (\ref{eqn2})  for $k_c= k = g$
plotted against $k$.}
 \label{fig4:01-03-20-gkdata.eps}
\end{figure}
%%%%%%%%%%%%%%%%%%%%%%%%%%%%%%%%%%%%%%%%%%%%%%%%%%%%%%%%%%

%%%%%%%%%%%%%%%%%%%%%%%%%%%%%%%%%%%%%%%%%%%%%%%%%%%%%%%%%% fig5
\begin{figure}[!h]
\centering
\includegraphics[width=10cm]{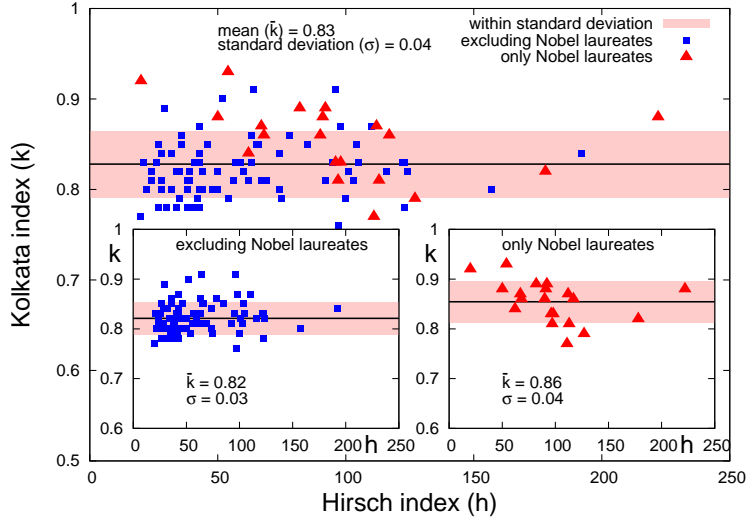}
 \caption{ Plot of the estimated values of Kolkata
index $k$ against Hirsch index $h$ \cite{Hirsch2005index} of 100
individual scientists, including 20 Nobel Laureates, from
the citation data  of Google  Scholar (Table \ref{table1}).
The separate insets clearly show that the
average value of $k$ index for Nobel Laureates ($k=0.86$) is distinctly higher than that ($k=0.83$) of  the scientists other than Nobel  Laureates.}
 \label{fig5:27jan-authors-h-k.eps}
\end{figure}
%%%%%%%%%%%%%%%%%%%%%%%%%%%%%%%%%%%%%%%%%%%%%%%%%%%%%%%%%%

%%%%%%%%%%%%%%%%%%%%%%%%%%%%%%%%%%%%%%%%%%%%%%%%%%%%%%%%%% fig6
\begin{figure}[!h]
%\centering
\includegraphics[width=10cm]{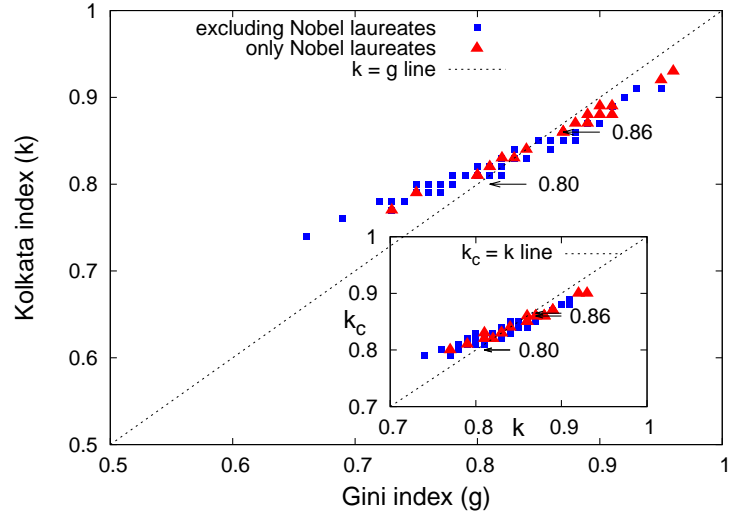}
 \caption{Plot of $k$ against $g$ for the citation
statistics of individual scientists from Table \ref{table1}.
It gives $k = 0.86\pm0.06$. This plot may be
compared with similar plot in Fig. \ref{fig4:01-03-20-gkdata.eps} for  paper
citation statistics of the universities,
institutions or journals.}
 \label{fig6:27jan-authors-g-k.eps}
\end{figure}
%%%%%%%%%%%%%%%%%%%%%%%%%%%%%%%%%%%%%%%%%%%%%%%%%%%%%%%%%%

\section{Data analysis for citations of papers by institutions and individuals}
\noindent First we  reanalyze the Web of Science data
\cite{Ghosh2014Inequality,Chatterjee2017Socio-economic} for the citations received by papers
published by scientists from a few selected
Universities and Institutions of the world
and citations received by papers published in
some selected Journals. We also added the
the analysis of the data from various World
Peace Organizations and Institutions \cite{Sinha2019Inequality}
for human deaths in different wars and
social conflicts. In Fig. \ref{fig4:01-03-20-gkdata.eps}, we  plot the
estimated values of $k$ against $g$  for
citations received by papers published
by authors from different universities
or institutes  and also of the publications
in different journals, as well as from data
for deaths distributions in various social
conflicts. Noting (see Fig. \ref{fig2:01-03-20-n-vs-gk.eps}) that $k$ has
approximately a piece-wise linear relationship
with $g$ as
\begin{equation}
 k = 0.5 + C g, 
 \label{eqn2}
\end{equation}
with a constant $C$, we estimate the $C$
values from the data points in Fig. \ref{fig2:01-03-20-n-vs-gk.eps}, and
using that we make a linear extrapolation for
$k_c=k=g$ (see the inset). It may be mentioned
that this approximate  linear relationship is
only phenomenologically  observed and fits the
values of $g$ and $k$ in  their lower range
both for this analytic form of Lorenz function
(see Fig. \ref{fig3:01-03-20-k-vs-g.eps}) and also for the observed data
(see Fig. \ref{fig4:01-03-20-gkdata.eps}).

We have estimated here the  values of Kolkata
index $k$ against the respective Hirsch index
$h$ \cite{Hirsch2005index} for $100$ individual scientists, including
$20$ Nobel Laureates (each having more than $100$
papers/entries and minimum $h$ index value $20$,
in their, `e-mail-site-verified', Google Scholar
page) from the respective paper citations (Table \ref{table1}).
In Fig. \ref{fig5:27jan-authors-h-k.eps}, we plot the estimated values of $k$
against $h$ of all these $100$ individual
scientists. The statistics suggests the $k$
index value ($0.83\pm0.04$) to be independent
of the $h$ index value (in the range $20\leq h\leq
222$). One inset shows the $k$ values ($k = 0.82
\pm 0.03$) plotted against respective $h$ values
for 80 scientists who are not Nobel Laureates
and another exclusively for the $20$ Nobel
Laureates. This clearly shows  that the limiting
values of $k$ index for the  Nobel Laureates on
average are  higher ($k =0.86\pm0.04$).
In Fig. \ref{fig6:27jan-authors-g-k.eps}, we show the plot of $k$ against the
$g$ values of their publication statistics and the
inset shows the estimated $k_c$ values obtained
using eqn. (\ref{eqn2}) and solving for $k_c = k = g$.

\begin{figure}[!h]
\centering
\includegraphics[width=10cm]{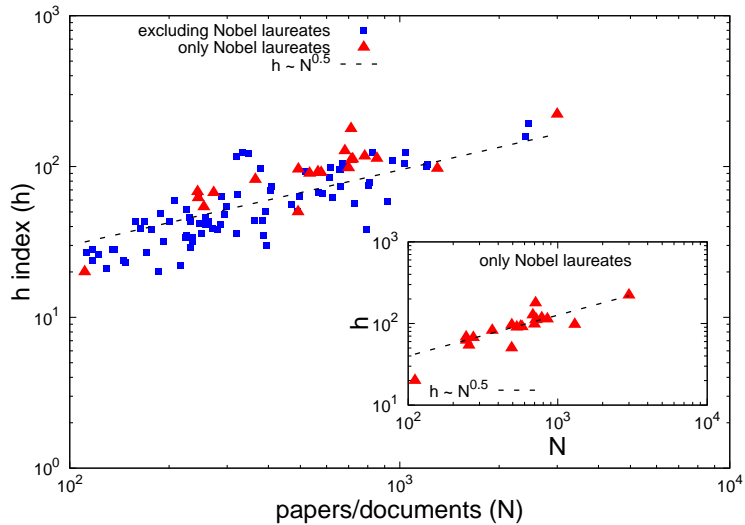}
 \caption{The data for Hirsch index $h$ values in Table \ref{table1}
suggest the relation $h \sim N^{\gamma}$, with
$N$ denoting total number of papers and
$\gamma \sim  0.5$.}
 \label{fig7:27jan-authors-n-h.eps}
\end{figure}

An interesting observation from Table \ref{table1} has been
that the $h$ index value of an author seems to
grow with number $N$ of publications,
statistically speaking, following a power law
$h \sim \sqrt N$ (see Fig. \ref{fig7:27jan-authors-n-h.eps}, where the inset for
the Nobel Laureates suggests a better fit).
%%%%%%%%%%%%%%%%%%%%%%%%%%%%%%%%%%%%%%%%%%%%%%%%%%%%%

\begin{table}[!htbp] %\small
%\small
\caption{Statistical analysis of the papers and their
citations for 100 `randomly chosen' scientists
(including 20 Nobel Laureates; denoted by *
before their names) in physics (Phys),
chemistry (Chem), biology/physiology/medicine
(Bio), mathematics (Maths),  economics
(Econ) and sociology (Soc), having  individual
Google Scholar page (with `verifiable email site')
and having at least 100 entries (papers or documents, latest not before 2018),
with Hirsch index ($h$) \cite{Hirsch2005index} value 20 or  above. These authors
(including the Nobel Laureates) have Hirsch index
 in the range 20-222 and number of papers
($N$) in the range 111-3000. The data were collected from Google Scholar during 1st week of January 2021 and names of the scientists appear here in the same form as in their respective Google Scholar pages.}
\label{table1}
\small
\begin{adjustwidth}{-0.85in}{-1in}% adjust the L and R margins by 1 inch
\begin{tabular}{|l|c|c|c|c|c|c|}
\hline
 name & total & total & \multicolumn{4}{c|}{index values}\\
 \cline{4-7}
%& paper & citations & Hirsch(h) & Gini (g) & kolkata & critical $k_c$\\
& paper & citations & $h$ & $g$ & $k$ & $k_c$\\
\hline
%\hline
*Joseph E. Stiglitz(Econ)&3000&323473&222&0.90&0.88&0.86\\
 \hline 
H. Eugene Stanley(Phys)&2458&200168&192&0.86&0.84&0.83\\
 \hline 
C. N. R. Rao(Chem)&2400&121756&157&0.77&0.80&0.81\\
 \hline 
*Hiroshi AMANO(Phys)&1300&44329&97&0.80&0.81&0.83\\
 \hline 
didier sornette(Phys)&1211&46294&103&0.80&0.81&0.82\\
\hline
Hans J. Herrmann-Phys)&1208&36633&100&0.75&0.79&0.81\\
\hline
Giorgio Parisi(Phys)&1043&88647&123&0.83&0.83&0.82\\
 \hline 
George Em Karniadakis(Math)&1030&53823&105&0.84&0.83&0.83\\
 \hline 
Richard G M Morris(Bio)&950&70976&110&0.89&0.87&0.85\\
 \hline 
debashis mukherjee(Chem)&920&15169&59&0.83&0.83&0.83\\
 \hline 
*Joachim Frank(Chem)&853&48077&113&0.80&0.81&0.82\\
 \hline 
R.I.M. Dunbar(Soc)&828&65917&124&0.81&0.82&0.82\\
 \hline 
 C. Tsallis(Phys)&810&36056&78&0.88&0.86&0.84\\
\hline
Biman Bagchi(Chem)&803&23956&75&0.77&0.79&0.81\\
 \hline 
Srinivasan Ramakrishnan(Phys)&794&6377&38&0.78&0.80&0.82\\
 \hline 
*William Nordhaus(Econ)&783&74369&117&0.87&0.86&0.85\\
\hline
Ronald Rousseau(Soc)&727&15962&57&0.83&0.83&0.82\\
\hline
*David Wineland(Phys)&720&63922&112&0.88&0.87&0.86\\
 \hline 
*Jean Pierre Sauvage(Chem)&713&57439&111&0.73&0.77&0.80\\
 \hline 
*Gregg L. Semenza(Bio)&712&156236&178&0.81&0.82&0.82\\
 \hline 
*Gérard Mourou(Phys)&700&49759&98&0.82&0.83&0.83\\
 \hline 
Jean Philippe Bouchaud(Phys)&688&44153&101&0.82&0.82&0.82\\
 \hline 
*Frances Arnold(Chem)&682&56101&127&0.75&0.79&0.81\\
 \hline 
Dirk Helbing(Phys)&670&60923&104&0.86&0.85&0.84\\
 \hline 
T. Padmanabhan(Phys)&662&26145&74&0.86&0.84&0.84\\
 \hline 
Gautam R. Desiraju(Chem)&661&59333&95&0.84&0.83&0.83\\
 \hline 
Brian Walker(Bio)&656&136565&96&0.93&0.91&0.89\\
 \hline 
A. K. Sood(Phys)&626&24076&62&0.82&0.81&0.81\\
 \hline 
Masahira Hattori(Bio)&618&80069&98&0.90&0.87&0.85\\
 \hline 
Joshua Winn(Phys)&611&45701&85&0.88&0.85&0.84\\
 \hline 
Kaushik Basu(Econ)&584&21506&66&0.86&0.85&0.84\\
 \hline 
*Abhijit Banerjee(Econ)&578&59704&91&0.89&0.88&0.86\\
 \hline 
Kimmo Kaski(Phys)&567&19647&67&0.80&0.81&0.82\\
 \hline 
*Esther Duflo(Econ)&565&69843&92&0.91&0.89&0.87\\
 \hline 
*Serge Haroche(Phys)&533&40034&90&0.87&0.86&0.85\\
 \hline 
Peter Scambler(Bio)&518&31174&92&0.81&0.81&0.82\\
 \hline 
Spencer J. Sherwin(Maths)&496&15383&63&0.83&0.83&0.83\\
 \hline 
*Michael Houghton(Bio)&493&49368&96&0.83&0.83&0.83\\
 \hline 
*A. B. McDonald(Phys)&492&20346&50&0.91&0.88&0.86\\
 \hline 
Mauro Gallegati(Econ)&491&10360&50&0.80&0.82&0.83\\
 \hline 
A. K. Raychaudhuri(Phys)&470&12501&56&0.78&0.81&0.82\\
 \hline 
Sidney Redner(Phys)&409&26287&74&0.78&0.80&0.81\\
 \hline 
Janos Kertesz(Phys)&407&20115&69&0.80&0.81&0.82\\
 \hline 
Jayanta K Bhattacharjee(Phys)&394&3674&30&0.74&0.78&0.81\\
\hline
Alex Hansen(Phys)&393&9678&50&0.76&0.80&0.82\\
\hline
Prabhat Mandal(Phys)&386&4780&35&0.75&0.79&0.81\\
\hline
Bikas K Chakrabarti(Phys)&384&10589&44&0.81&0.82&0.83\\
 \hline 
Ashoke Sen(Phys)&379&33342&97&0.69&0.76&0.80\\
 \hline 
*Paul Milgrom(Econ)&365&102043&82&0.90&0.89&0.87\\
 \hline 
Ramasesha S(Chem)&362&7188&44&0.78&0.80&0.82\\
 \hline 
%\hline
\end{tabular}
\begin{tabular}{|l|c|c|c|c|c|c|}
\hline
 name & total & total & \multicolumn{4}{c|}{index values}\\
 \cline{4-7}
%& paper & citations & Hirsch(h) & Gini (g) & kolkata & critical $k_c$\\
& paper & citations & $h$ & $g$ & $k$ & $k_c$\\
\hline 
Noboru Mizushima(Bio)&347&117866&122&0.82&0.83&0.83\\
\hline
William S. Lane(Bio)&334&72622&123&0.74&0.78&0.80\\
 \hline 
Debraj Ray(Econ)&322&23558&65&0.85&0.85&0.85\\
\hline
Beth Levine(Bio)&321&103480&116&0.81&0.82&0.83\\
\hline
Debashish Chowdhury(Phys)&320&8442&36&0.88&0.86&0.84\\
 \hline 
Toscani Giuseppe(Math)&299&10129&54&0.75&0.79&0.82\\
 \hline 
Matteo Marsili(Phys)&294&8976&48&0.77&0.80&0.82\\
 \hline 
Rosario Nunzio Mantegna(Phys)&289&29437&63&0.88&0.86&0.85\\
 \hline 
Diptiman Sen(Phys)&286&6054&41&0.74&0.78&0.80\\
 \hline 
J. Barkley Rosser(Econ)&281&5595&38&0.81&0.81&0.82\\
 \hline 
*David-Thouless(Phys)&273&47452&67&0.89&0.87&0.86\\
 \hline 
Sanjay Puri(Phys)&271&6053&39&0.79&0.81&0.82\\
\hline
Maitreesh Ghatak(Econ)&263&11942&43&0.89&0.87&0.86\\
 \hline 
Serge GALAM(Phys)&258&7774&41&0.82&0.83&0.84\\
 \hline 
Sriram Ramaswamy(Phys)&257&13122&46&0.87&0.85&0.84\\
 \hline 
*Paul Romer(Econ)&255&95402&54&0.96&0.93&0.90\\
\hline
Krishnendu Sengupta(Phys)&251&7077&36&0.86&0.85&0.84\\
\hline
Chandan Dasgupta(Phys)&248&6685&42&0.76&0.79&0.81\\
 \hline 
Scott Kirkpatrick(CompSc)&245&80300&64&0.95&0.91&0.88\\
 \hline 
*richard henderson(Chem)&245&27558&62&0.84&0.84&0.84\\
 \hline 
*F.D.M. Haldane(Phys)&244&41591&68&0.87&0.86&0.86\\
 \hline 
Kalobaran Maiti(Phys)&235&3811&32&0.86&0.84&0.83\\
 \hline 
Amitava Raychaudhuri(Phys)&235&3522&34&0.74&0.78&0.81\\
 \hline 
Bhaskar Dutta(Econ)&232&6945&43&0.82&0.83&0.84\\
 \hline 
Ganapathy Baskaran(Phys)&232&6863&29&0.91&0.89&0.87\\
\hline
Hulikal Krishnamurthy(Phys)&231&14542&46&0.86&0.85&0.84\\
 \hline 
Rahul PANDIT(Phys)&226&6067&35&0.82&0.82&0.82\\
\hline
W. Brian Arthur(Econ)&225&47014&52&0.92&0.90&0.88\\
 \hline 
Pratap Raychaudhuri(Phys)&224&4231&34&0.80&0.82&0.83\\
 \hline 
Jose Roberto Iglesias(Phys)&217&1819&22&0.77&0.80&0.82\\
 \hline 
%B. Douglas Bernheim(Econ)&209&40973&73&0.81&0.83&0.85\\
%\hline
Hongkui Zeng(Bio)&208&18914&60&0.82&0.82&0.83\\
\hline
Deepak Dhar(Phys)&200&7401&43&0.77&0.80&0.82\\
 \hline 
Sitabhra Sinha(Phys)&193&2855&32&0.76&0.80&0.83\\
 \hline 
Amol Dighe(Phys)&189&8209&49&0.76&0.80&0.82\\
 \hline 
Arup Bose(Maths)&186&1965&20&0.73&0.77&0.79\\
 \hline 
Abhishek Dhar(Phys)&177&5004&38&0.73&0.78&0.80\\
\hline
S. M. Bhattacharjee(Phys)&171&2268&27&0.72&0.78&0.81\\
 \hline 
Martin R. Maxey(Maths)&168&10124&43&0.86&0.84&0.83\\
 \hline 
Arnab Rai Choudhuri(Phys)&164 &6115&39&0.81&0.82&0.83\\
 \hline 
Victor M. Yakovenko(Phys)&158&7699&43&0.72&0.78&0.81\\
 \hline 
Md Kamrul Hasan(Phys)&147&1844&23&0.66&0.74&0.79\\
 \hline 
 Shankar Prasad Das(Phys)&145&2476&24&0.81&0.81&0.81\\
 \hline
Amit Dutta(Phys)&137&2845&28&0.79&0.81&0.82\\
 \hline 
Anirban Chakraborti(Phys)&135&4809&28&0.83&0.84&0.85\\
 \hline 
Parongama Sen(Phys)&129&3062&21&0.82&0.83&0.83\\
 \hline 
Roop Mallik(Bio)&122&3363&26&0.83&0.83&0.84\\
\hline
Wataru Souma(Phys)&117&2607&24&0.82&0.82&0.83\\
 \hline 
Subhrangshu S Manna(Phys)&117&4287&28&0.75&0.80&0.83\\
 \hline 
Damien Challet(Math)&112&5521&27&0.86&0.85&0.85\\
 \hline 
*Donna Strickland(Phys)&111&10370&20&0.95&0.92&0.90\\
\hline
%Chandrima-Das-BIo)&104&2473&22&0.81&0.82&0.83\\
%\hline
\end{tabular}
\end{adjustwidth}
\end{table}

%%%%%%%%%%%%%%%%%%%%%%%%%%%%%%%%%%%%%%%%%%%%%%%%%%%%%%%

%%%%%%%%%%%%%%%%%%%%%%%%%%%%%%%%%%%%%%%%%%%%%%%%%%%%%%%%%% fig7
%%%%%%%%%%%%%%%%%%%%%%%%%%%%%%%%%%%%%%%%%%%%%%%%%%%%%%%%%%

%limiting  value (as a universal social number).

%Google Scholar

\section{Summary and discussions}
\noindent Social inequalities in every aspects, resulting
from competitiveness are described by various
distributions (like Log-normal, Gamma, Pareto,
etc., see e.g., \cite{Chakrabarti2013Income,Sen2014Sociophysics}). Economic inequality
has long been characterized  \cite{Gini1921Measurement} by the Gini
index ($g$) and a few other (much) less popular
geometric characterizations (see e.g., \cite{Eliazar2015sociogeometry}) of
the Lorenz curve or function $L(x)$ \cite{Lorenz1905Methods} (see
Fig. \ref{Lorentz2.pdf}). We introduced \cite{Ghosh2014Inequality} the Kolkata index
$(k)$ as a fixed point of the Complementary
Lorenz function $L^{(c)}(x) (L(x)$ has trivial
fixed points at $x$ = 0 and 1). In fact, the
Kolkata index $k$ is also (geometrically) related to
the `perpendicular diameter' \cite{Eliazar2015sociogeometry,Eliazar2016Harnessing} of the Lorenz
curve. Unlike the Gini index, which measures some
average properties
of the Lorenz function, Kolkata index gives a
tangible interpretation: $(1 - k)$ fraction of
rich people or papers or social conflicts
possess or attract or cause $k$ fraction of
wealth or citations or deaths respectively.

Assuming a generic form $L(x) = x^n$ (as
in eqn. (\ref{eqn1}), giving $L(0) = 0$ and $L(1) = 1$
and monotonic increase parametrized by $n$, as
desired), we see (in Figs. \ref{fig2:01-03-20-n-vs-gk.eps} and \ref{fig3:01-03-20-k-vs-g.eps}) that as
inequality increases (with increasing $n$) from
equality $k = 0.5$ and $g = 0$ for $n = 1$ to
extreme inequality $k=g=1$ as $n \to \infty$,
$k$ has a non-monotonic variation with
respect to $g$ such that $k$ and $g$ crosses
at $k=g \simeq 0.86$ and they finally meet at
$k=g=1$. As the Gini index $(g)$ is
identified (see \cite{Biro2020Gintropy}) as the information entropy of social systems and the Kolkata
index ($k$) as the inverse of effective temperature
of such systems (increasing $k$ means decreasing
average money in circulation and hence decreasing
temperature \cite{Chakrabarti2013Income}), this multivaluedness of
(free energy) $g/k$ as function of (temperature)
$k^{-1}$ at $g = k \simeq 0.86$  and $g = k = 1$
(Figs. \ref{fig2:01-03-20-n-vs-gk.eps} and \ref{fig3:01-03-20-k-vs-g.eps}) indicates a first order like
(thermodynamic) phase transition \cite{Stanley1971intro} at $g = k \simeq
0.86$.

We also noted \cite{Ghosh2014Inequality,Ghosh2016kinetic,Chatterjee2017Socio-economic,Sinha2019Inequality,Banerjee2020Kolkata1,Banerjee2020Kolkata2} that the $k$ index value, in extreme limits of social
competitiveness, converge towards a high value
around $0.86\pm0.03$ (see Fig. \ref{fig4:01-03-20-gkdata.eps}), though
not near the highest possible value $k = 1$
(maximum possible value for extreme inequality).
Indeed, $k$ index gives a quantitative
generalization of the century old $80-20$ rule
($k=0.80$) of Pareto \cite{Pareto1971Translation} for economic inequality
(though, as mentioned earlier, the economic
inequality statistics today for various
countries of the world shows \cite{Ghosh2014Inequality} much lower
$k$ values in the range $0.61-0.73$, because
of various economic welfare measures).

In summary, using a generic form (valid for all kinds
of inequality distributions) of the Lorenz
function $L(x)$ (= 0 for $x = 0$  and = 1
for $x = 1$ and monotonically increasing
in-between), we showed (see Fig. \ref{fig2:01-03-20-n-vs-gk.eps}) that as
inequality increases (with increasing
values of $n$  in eqn. (\ref{eqn1})), the difference
in values between $k$ (initially higher in
magnitude) and $g$, both individually increasing, vanishes at $k = g \simeq 0.86$
(after which $g$ becomes higher than $k$ in
magnitude) until the point of extreme
inequality ($n \to \infty$) where $k = 1
= g$, where they touch each other in
magnitude. We consider this crossing point
of $k = g \simeq 0.86$, which is higher than
the Pareto value $0.80$ \cite{Pareto1971Translation}), as an attractive stable fixed point inducing a saturation and universal value
for the inequality measure $k$ for the various
distributions in different social sectors. Indeed,
this limiting universal value of $k$ may
effectively restrict the range of interactions
among the agents, depending on the dynamic
interplay between them in the socio-dynamical
models (see e.g., \cite{Sen2014Sociophysics,Solomon2001Power}). This saturation value of the $k$ index also restrict the exponent value of the power law distribution for income, wealth etc. (see \cite{Ghosh2016kinetic,Solomon2001Power}).

Our earlier citation analysis \cite{Ghosh2014Inequality,Chatterjee2017Socio-economic}  from
the web of science data for citations against
papers published by authors from different
established universities or institutes and
also of the publications in different
competitive journals indicate the limiting
value of $k$ to be $0.83\pm0.03$. Similar
analysis  for human deaths in different
deadly wars or social conflicts \cite{Sinha2019Inequality} also
suggests similar limiting value of $k$
(see Fig. \ref{fig4:01-03-20-gkdata.eps}). These are a little higher
than the Pareto value \cite{Pareto1971Translation} of $k$ ($=0.80$).
It may be noted that, unlike the economic
welfare measures taken to avoid social unrest
(revolutions in earlier era and strikes etc
these days), the universities and institutions
encourage competence and discourage failure.
Competition in the wars etc are of course
extremely fierce.

%%%%%%%%%%%%%%%%
Our citation analysis here of $100$ individual
scientists, including $20$ Nobel Laureates
(see Table \ref{table1}), in different scientific and
sociological subjects (each having at least
$100$ papers or entries $N$ in their
respective Google Scholar page, with
`verifiable email site’, and having the
Hirsch index $h$ value $20$ or more) suggests
$k = 0.83\pm0.04$ (see Fig. \ref{fig5:27jan-authors-h-k.eps}) and independent
of the $h$ index value in the range $20-222$.
Indeed, for Nobel Laureates, the the average
value of the Kolkata index is slightly higher
($k =0.86\pm0.04$, again independent $h$ index
value) saying that for any of them, typically about
$14\%$ of their successful papers earn  about $86\%$
of their total citations. It may be interesting
to note that Google Scholar has developed a page \cite{web1}
on Karl Marx, father of socialism, and it
is maintained by the British National Library.
The page contains $3000$ entries (books, papers,
documents, published by Marx himself, or later
collected, translated, and edited by other individuals
and different institutions. According to this page, his
total citation counts $387995$ and his Hirsch index
value ($h$) is $213$. Our citation analysis says, his
Kolkata index value ($k$) is $0.88$, suggesting
again that $88\%$ of his citations comes from
$12\%$ of his writings! From Table \ref{table1}, we also
note that individual’s $h$ index value grows,
on average, with the total number $N$ of his/her
publications as $h \sim \sqrt  N$ (more clearly
so for the Nobel Laureates; see the inset of Fig. \ref{fig7:27jan-authors-n-h.eps}), and
as such Hirsch index has no saturation value
(as a universal limiting  social number). It
may be mentioned in this connection that our
observation regarding the growth of Hirsch
index value with the volume or number of
publications by individual authors seem to
suggest a similarity with the Heaps' law \cite{Heaps}
in linguistics, where the number of distinct
words in a document grows following a power
law with the  document size (having exponent
value in the range $0.4-0.6$).

%%%%%%%%%%%%%%%%

To conclude, based on  our analytic study of the properties of the Gini ($g$) and Kolkata ($k$)
indices for social inequality, based on a
generic form (eqn. (\ref{eqn1})) of the Lorenz function $L(x)$
(in section II), and some observations on the
citation statistics of individual authors
(including Nobel laureates), institutions and
journals (also on the statistics of deaths in
wars etc), we make a conjecture that about
$14\%$ of people or papers or social conflicts
earn or attract or cause about $86\%$ of wealth
or citations or deaths in very competitive
situations in the markets, universities or wars respectively.
This is a modified form of the (more than a)
century old $80-20$ law of Pareto in economy
which is not visible in today's economies because
of various welfare strategies to mitigate poverty).
This limiting value of the  Kolkata index for
inequality $k(\simeq 0.86)$ gives perhaps
an universal social constant or number.

\section*{acknowledgments}
\noindent BKC is grateful to INSA Senior Scientist
grant (from Indian National Science Academy)
for support. We are extremely grateful to
Soumyajyoti Biswas, Indrani Bose, Anirban
Chakraborti, Arnab Chatterjee and
Manipushpak Mitra for useful comments on
the manuscript. We are also thankful
to the two anonymous referees for raising some
thoughtful points and  suggestions.

\end{document}